\documentclass[aps,floatfix,superscriptaddress,notitlepage,nofootinbib]{revtex4-1}
\usepackage{amsmath,amssymb,amsfonts,graphics,graphicx,dcolumn,bm,enumerate}
\usepackage{comment,natbib,appendix}
\usepackage{multirow,color}
\usepackage{chngpage}
\usepackage{afterpage}
\usepackage{xcolor}
\usepackage{amsthm}
\usepackage{natbib}
\usepackage{hyperref}
\usepackage[margin=1.2in]{geometry}
\usepackage{epstopdf}
\usepackage{float}

\newcommand{\cmp}
{\affiliation{Saha Institute of Nuclear Physics, Kolkata 700064, India.}}
\newcommand{\isi}
{\affiliation{Economic Research Unit, Indian Statistical Institute, Kolkata 700108, India.}}
\newcommand{\raghunathpur}
{\affiliation{Department of Physics, Raghunathpur College, Raghunathpur, Purulia 723133, India.}}

\newcommand{\Vidyasagar}
{\affiliation{Department of Physics, Vidyasagar College, Kolkata 700006, India.}}

\begin{document}
\title{Kinetic Models of Wealth Distribution Having
Extreme Inequality:  Numerical Study of Their
Stability Against Random Exchanges}
\author{Asim Ghosh}
\email[Corresponding author: Email: ]{asimghosh066@gmail.com}
\raghunathpur

\author{Suchismita Banerjee}
\isi

\author{Sanchari Goswami}
\Vidyasagar

\author{Manipushpak  Mitra}
\isi
 \author{Bikas K. Chakrabarti }%
 \isi \cmp 
 
\begin{abstract}
In view of some persistent recent reports on a singular kind of growth
of the world wealth inequality, where a finite (often handful) number
of people tend to possess more than the wealth of the planet's 50\% population,
we explore here if the kinetic exchange models of the market can ever
capture such features where a significant fraction of wealth can
concentrate in the hands of a countable few  when the market size $N$
tends to infinity. One already existing example of such a kinetic
exchange model is the Chakraborti or Yard-Sale model, where (in
absence of tax redistribution etc) the entire wealth condenses in the
hand of one (for any value of $N$), and the market dynamics stops. With tax
redistribution etc, its steady state dynamics have been shown to have
remarkable applicability in many cases of our extremely unequal
world.  We show here that another kinetic exchange model (called here the
Banerjee model) has intriguing intrinsic dynamics, by which
only ten rich traders or agents possess about 99.98\% of the total
wealth in the steady state (without any tax etc like external manipulation)
for any large value of $N$. We will discuss in some detail the statistical
features of this model using Monte Carlo simulations. We will also show,
if the traders each have a non-vanishing  probability  $f$ of following
random exchanges, then these condensations of wealth (100\%  in the
hand of one agent in the Chakraborti model, or about 99.98\% in the hands ten agents
in the Banerjee model) disappear in the large $N$ limit. We will also
see that due to the built-in possibility of random exchange dynamics
in the earlier proposed Goswami-Sen model,  where the exchange
probability decreases with an inverse power of the wealth difference
of the pair of traders, one did not see any wealth condensation
phenomena. These aspects of the statistics of these
intriguing models have been discussed
here.
\end{abstract}

\maketitle

\section{INTRODUCTION}
The first successful theory of classical many-body
physics or classical condensed matter systems has
been about one and a quarter centuries old kinetic theory
of the (classical) ideal gas, composed of Avogadro number
(about $10^{23}$) order constituent atoms or molecule
(each following Newtonian dynamics). It remains still
a robust, versatile and extremely successful foundation
of classical many-body physics. Social systems, economic
markets in particular, are intrinsically many-body dynamical
systems composed of a lesser number of constituents (order of $10^2$
for a village market to the order of
$10^{10}$ for a global market). One Robinson Crusoe in
an island can not develop a  market or a society for that
matter and markets are intrinsically many-body systems. It is no wonder that the kinetic exchange of money
or wealth models had therefore been conjectured early on (e.g., by
Saha and 
 Srivastava \cite{Saha1931} in 1931, Mandelbrot \cite{Mandelbrot1960} in
1960) and resurrected recently (e.g., by
Chakrabarti and Marjit \cite{Chakrabarti1995} in 1995, Dragulecu
and Yakovenko \cite{Dragulescu2000} in 2000, Chakraborti and
Chakrabarti \cite{Chakraborti2000} in 2000, Chatterjee, Chakrabarti
and Manna \cite{Chatterjee2004a} in 2004).

The kinetic exchange models of trades and their
statistics have been quite successful in capturing
several realistic features of wealth distributions
among the agents in the societies (see e,g.,
\cite{Yakovenko2009,Chakrabarti2013}). The beneficial effects of the agent's
saving propensity in reducing the social
inequality has been studied extensively \cite{Chakraborti2000,Chatterjee2004a,Chakrabarti2013}.
The choice of the poorest trader as mandatory in
each trade (the other trade partner being randomly
chosen) leads to the remarkable self-organized
poverty line, below which none remains in the steady
state (see e.g., \cite{Pianegonda2004,Iglesias2010,Ghosh2011,Paul2022}). This model was inspired
by some crucial observations by the economists
(see e.g., \cite{Iglesias2010}) and suggests the built-in (self-organized) remedies for reducing social inequality.
Though it must be admitted, such intriguing
self-organizing properties of the kinetic exchange
models have not been investigated extensively yet.

Contrarily, the recent focus has moved to the
unusual rate of growth of social inequality in
the post world war II period (see e.g., \cite{Pickety2014,Iglesias2021,Danial2022,Banerjee2023}),
which in some countries seem to have crossed
significantly above the 80-20 Pareto limit and
have reached a steady state with 87\% percent
wealth accumulated in the hands of 13\% people.
This has indeed been argued, following an analogy with the
inequality index values for the
avalanche burst statistics in
self-organized sand-pile models near their
respective critical points, to be the natural
limit in all social competitive situations,
where the welfare mechanisms (helping those who
fail to participate properly in such
self-organizing dynamics) are either absent or
removed (see e.g., a recent review \cite{Banerjee2023}).

Although the Pareto-like inequality mentioned
above where a small fraction of people (say 13\%)
possess a large fraction (say 87\%) of  wealth,
can already be devastating, more annoying kind
of inequalities are being reported recently. For
example, the Oxfam Report \cite{Oxfam2020} of January, 2020
in Davos said ``The world’s 2,153 billionaires
have more wealth than the 4.6 billion people who
makeup 60 percent of the planet’s population."
In other words, a handful number (about $10^3$)
of  rich people possess more than  about 60\%
(or $10^9$ order) poor people's  wealth of this
planet.  This dangerous trend in the world as
a whole, is repeatedly mentioned in
various recent reports.

The Pareto-type inequality mentioned above have
long been investigated (see e.g., \cite{Chatterjee2004a,Chatterjee2007})
using the kinetic exchange models with non-uniform
saving propensities of the traders (see e.g., \cite{Chakrabarti2013},
\cite{Pareschi2014} for reviews).  One may naturally ask 
the question, does the kinetic exchange theory  allow
for possible models where only a handful of traders
(say, about 10 in number) possess a significant
fraction (say, above 50\%) of the total wealth
considered in the model, even when its population
$N$ tends to infinity?

The answer is yes.  The Chakraborti model \cite{Chakraborti2002},
widely known today as the Yard-Sale model, starting
with \cite{Hayes2002}, have attracted a lot of attention (see
e.g., \cite{Cardoso2023,Boghosian2019,Julian2022}). In its barest form \cite{Chakraborti2002}  in the Chakraborti model (called C-model
here),  two randomly chosen traders at
any point of time come and participate in an exchange
trade when the richer  one saves the excess over the
poorer's wealth and goes for a random exchange of
the total available wealth (double of the poorer's).
The slow but inevitable attractor fixed point of
the trade dynamics arrive when all wealth ends up
in the hand of just one trader, no matter how big
the population ($N$) is. Because of the particular
form of savings during any trade, whenever one becomes
pauper, nobody trades with him and gradually all
condense to that state where one trader
acquires  the entire wealth and the trade dynamics
stop (see also \cite{Cardoso2023}). External perturbations like
regular redistribution of tax collections  by the
central government (or any non-playing agent) can
help relieving \cite{Boghosian2019,Julian2022} the condensation phenomenon
and this seems to fit well with  many observed
situations \cite{Boghosian2019}. We will  show here numerically that
if each of the traders has  a finite probability ($f$)
of playing Dragulecu and Yakovenko (DY) \cite{Dragulescu2000} type
random exchanges, then for any $f > 0$, the
condensation  of wealth in the hand of one trader
disappear and the steady state distribution of
wealth becomes exponentially decreasing, as in the
DY model.

In the Goswami-Sen (or GS) model \cite{Goswami2014}, one considers a
kinetic exchange mechanism where the interaction (trade)
probability among the trade partners ($i$ and $j$)
decreases with their wealth difference ($|m_i - m_j|$)
at that instant of trading (time), following a power
law ($|m_i - m_j|^{-\alpha}$). Of course, for $\alpha$
= 0, the model reduces to that of DY. Their numerical results showed that  for $\alpha$ values less than
about 2.0, the  steady-state wealth distribution among the  traders
are still DY-like (exponentially  decaying with increasing wealth).
For higher values (beyond 2.0) of $\alpha$, power-law  
(Pareto-law) decays occur. No condensation of wealth in the hands of a
finite number of traders or agents are observed, because of the
inherent DY-like exchange probability in the dynamics of the model (checked by extrapolating with $N$ the fraction
of total wealth possessed in the steady state by the richest ten
traders).

We finally consider here a seemingly natural version of the
kinetic exchange model, called here the Banerjee (B) model \cite{Banerjee2021}, where
the intrinsic dynamics of the model lead to  another extreme kind of inequality in the
steady state in the sense that precisely ten traders (out of the $N$
traders in the market;  $N \rightarrow \infty$) possess  (99.98 $\pm$
0.01)\% of the total wealth.  
These fortunate traders are not unique
and their fortune does not last for long
(residence time on average is about 66 time
units with the most probable value around
25 time units, counted in units of $N$
trades or exchanges, for any value of $N$)
and it decreases continuously with
increasing fraction ($f$) of random trades
or interactions.
 Unlike in the Chakraborti or
Yard-Sale model \cite{Chakraborti2002,Hayes2002}, where the dynamics stops
after the entire wealth goes to one (unless perturbed externally), here
the trade dynamics continue with the total wealth circulating only
within a handful (about ten) traders in the steady state. In this
model, after each trade, the traders are ordered from lowest wealth to
highest and each of the traders trade only with their
nearest-in-wealth traders, richer or poorer compared to own, with
equal probability. Even if by chance the entire wealth goes to one
trader, the dynamics of random exchanges do not stop in this model as
all the paupers become the only nearest neighbors (wealth-wise) of
this trader and random exchange among them occurs! The process
continues. Apart from the steady state wealth distributions and most
probable wealth amounts of the top few rich traders, we will
show that in this model the condensation of almost the entire (99.98\%)
wealth occurs in the hands of 10 traders (no matter how big $N$ is).
We will show here again that this condensation disappears when a
finite fraction $f$ of time each of the traders go for DY type random
exchanges, and eventually the DY-type exponentially decaying wealth
distribution sets-in, after a power law region for low values of $f$.

\section{MODELS and NUMERICAL STUDIES FOR THEIR STATISTICS}
\begin{figure}[!tbh]
    \centering
    \includegraphics[scale=0.75]{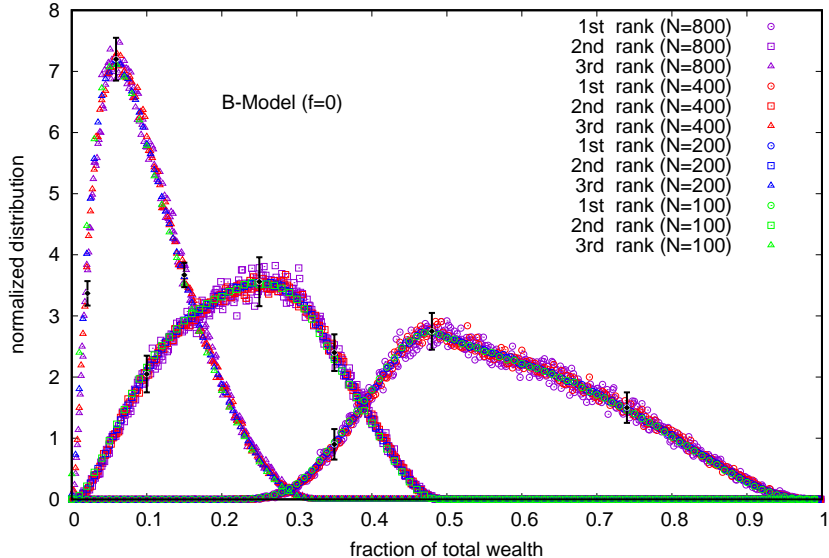}
    \caption{Distributions  of the fraction of total wealth ($M = N$) ending up in
the hands of the richest three traders. The error estimation is based on 10 runs. The typical errors in the distribution of
wealth are seen to grow with $N$ near the
most-probable value of the wealth fraction
and are  indicated for $N$ = 800 for all
the three traders. Far away from the
most-probable values, the errors are less
than the data point symbol sizes.}
    \label{fig:1}
\end{figure}
We study here numerically the statistical features of the  three
kinetic exchange models introduced in the Introduction. We begin with
the B (Banerjee  \cite{Banerjee2021}) model. Next, we consider the  C  (Chakraborti, or
Yard Sale) model \cite{Chakraborti2002,Hayes2002} and then the GS (Goswami-Sen)  model \cite{Goswami2014}.
In order to explore the stability
of the condensation of wealth feature in these models, we study the
steady-state wealth distributions  $P(m)$ in each of these models and
the fraction of total wealth concentrated in the hands of a few (say
ten) traders or agents (whenever meaningful), allowing each trader to
have a nonvanishing probability $f$ (the faction of tradings or times) to
go for DY (Dragulecu and Yakovenko \cite{Dragulescu2000}) type random exchanges.

Most of the numerical (Monte Carlo) studies of the dynamics of these
models are studied with four sets of numbers $N$ of the agents or
traders: $N$ = 100, 200, 400, and 800 and at each time step $t$, the
dynamics runs over all the $N$ order traders. We consider total money
($M$), to be distributed among the agents is equal to $N$ and we
denote the money of any agent  $i$ at time $t$ by $m_i(t)$ and, as
such $M = \sum_i m_i(t) = N$.  When the steady state is reached after
the respective relaxation times  when the average quantities do not
change with time (relaxation time typically much less
than $10^5$ trades/interactions for the $N$
values considered here),  the statistical quantities  are evaluated from
averages over $10^5$ post-relaxation time steps or so.

\subsection{Banerjee model results}
\begin{figure}[!tbh]
    \centering
    \includegraphics[scale=0.75]{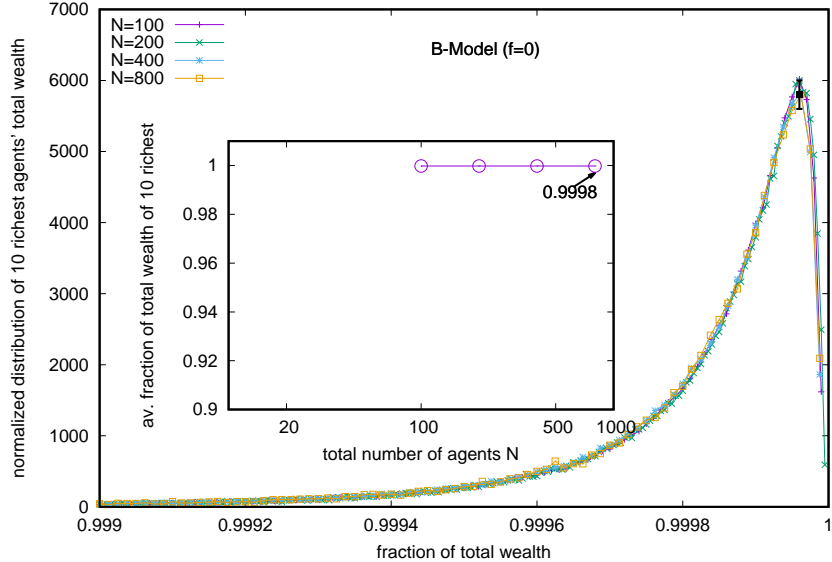}
    \caption{Distribution of the total wealth fraction  possessed by the
ten richest  (at any time in the steady state and for different $N$
values). The inset shows that the average of this total wealth
fraction of the ten richest (for any time and any value of $N$) 
  in the steady state is very close to 0.9998.
It may be noted, although the wealth share fractions
of the richest ten traders have considerable
fluctuations (see Fig. \ref{fig:1}), the sum total of their
wealth fractions have hardly any fluctuation (much
less than the symbol size in the inset). The error estimation is based on 10 runs. The
typical error in the distribution of
total wealth of the ten richest are seen
to be more than the data point  symbol
sizes only near the most-probable value,
where it is indicated.}
    \label{fig:2}
\end{figure}

\begin{figure}[!tbh]
    \centering
    \includegraphics[scale=0.75]{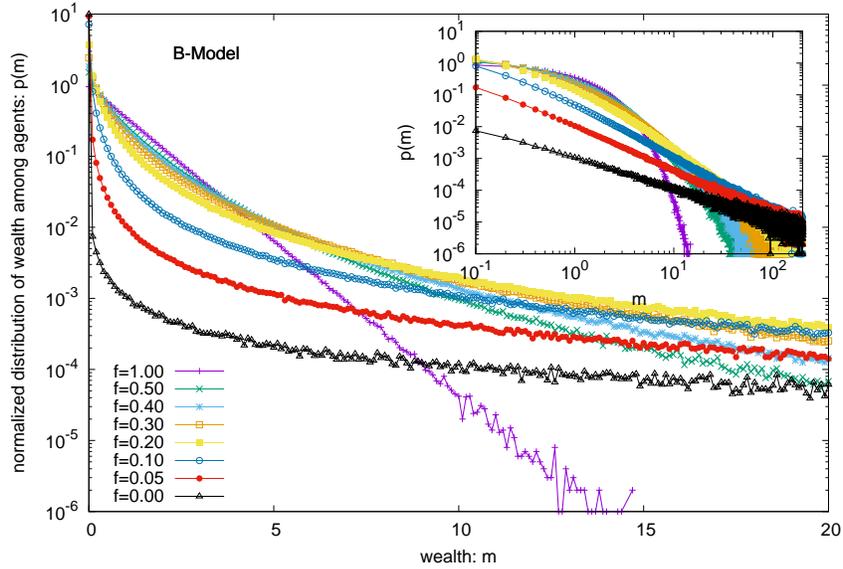}
    \caption{Wealth distribution $P(m)$ among all the agents against the
wealth $m$ in the B model for different probability  $f$ of DY random
exchanges. Note that
the fluctuations appear to grow more for the
lower values of the distribution of wealth
due to the log scale used in the y-axis.}
    \label{fig:3}
\end{figure}

In this B-model, when the DY fraction ($f$)
is set equal to zero, no wealth distribution
$P(m)$ across the population is meaningful,
because of wealth condensation in the hands
of a few.  We first study the distributions
(see Fig. \ref{fig:1}) of total wealth fraction in the
hands of the richest three. Note, these three
are not unique, and once they become so rich,
their residence time (in unit of $N$) is
finite (about 66) and in case these positions
are lost, the return time also is finite.

Though the distributions  of the total wealth fraction in the  hands
of  the few richest (shown in Fig. \ref{fig:1}) are rather wide (each one spread
over more than 30\% of the total wealth and does not $N$), the
distribution of the total wealth fraction  possessed by the ten
richest  (at any time in the steady state) is extremely narrow and
spreads over 0.1\% only (see Fig. \ref{fig:2}). At any  time in the steady state
its value is much more robust in this B model (with $f$ = 0) and its
value is less than unity, but very close to 0.9998.

\begin{figure}[!tbh]
    \centering
    \includegraphics[scale=0.75]{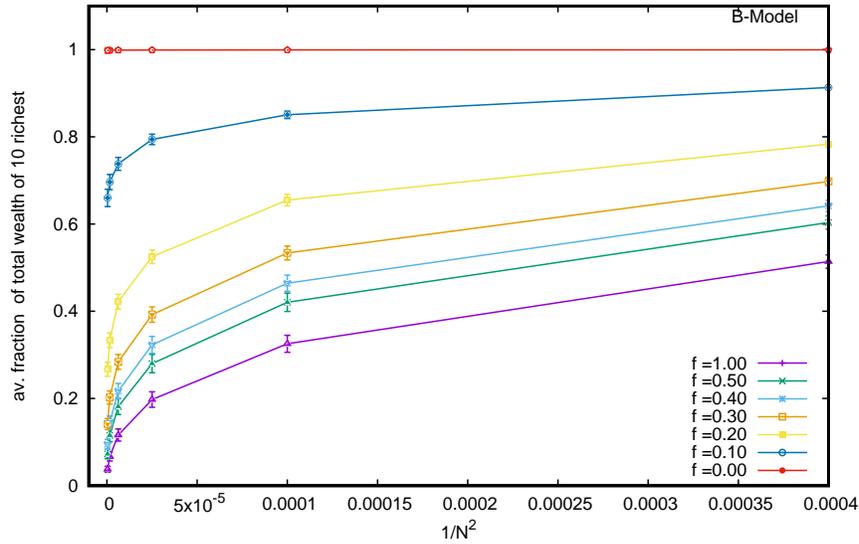}
    \caption{To get the limiting values (for large $N$) of the
average fraction of total wealth ($M = N$) possessed by the ten
richest traders in the steady state, we plot the fraction against
$1/N^2$ (as with DY type trades each of N traders interacts with (N -
1 other trader).  The extrapolated values  all seem to approach
zero for any non-zero value of $f$ (but remains a constant 0.9998 for
$f$ = 0, as in  the pure B model). The error estimation is based on 10 runs. Typical sizes of error bars are indicated.}
    \label{fig:4}
\end{figure}

Next,  we consider the B model with a nonvanishing probability $f$ of
each trader to follow DY trades or exchanges. We see, immediately, the
wealth condensation disappears and with increasing values of $f$, the
wealth gets Boltzmann (exponentially) distributed among all the agents
(see Fig. \ref{fig:3}), starting with Pareto-like power law distribution for
lower values of $f$ (see the inset of Fig. \ref{fig:3}).  Indeed, when we
consider the limiting values (for large $N$) of the average fraction
of total wealth ($M = N$) possessed by the ten richest traders in the steady state, they all seem to vanish (see Fig. \ref{fig:4}) for any non-zero value of
$f$ (and remains a constant 0.9998 for $f$ = 0, the pure
B model).

\begin{figure}[!tbh]
    \centering
    \includegraphics[scale=0.6]{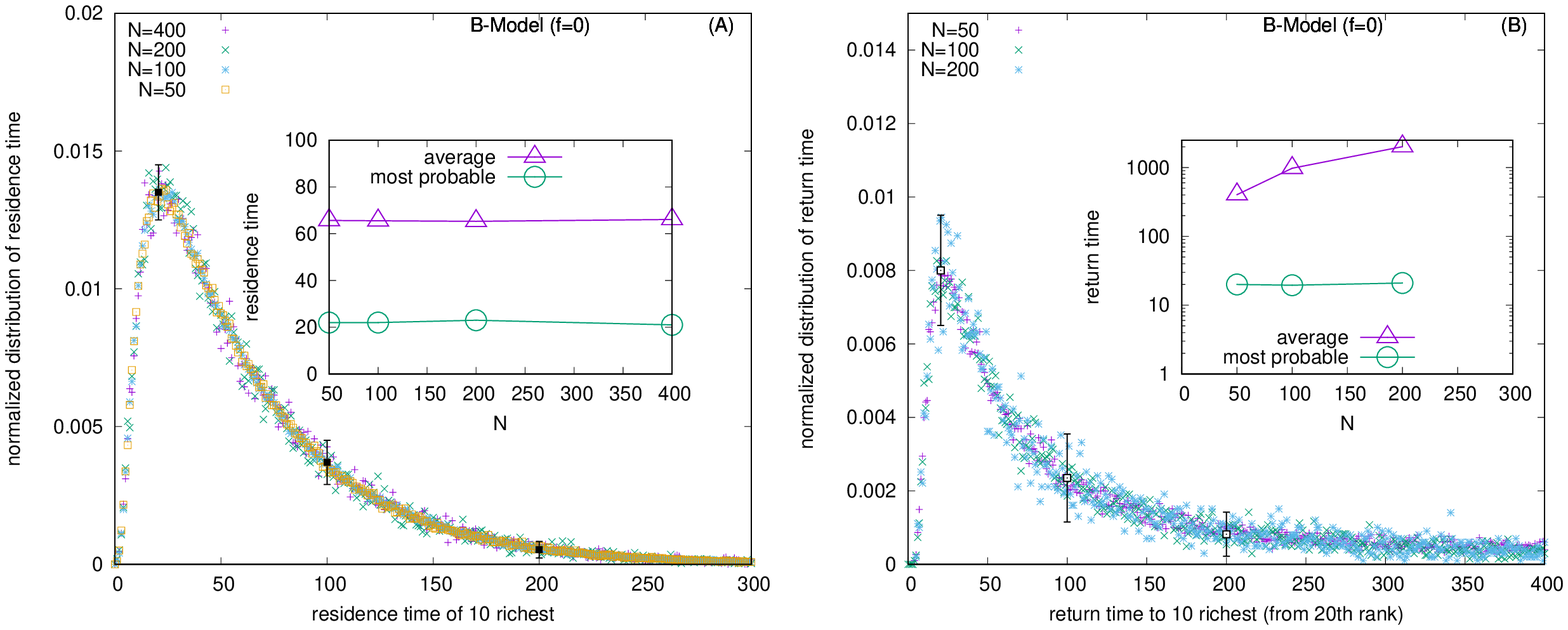}
    \caption{
(a) The distribution of residence-time (in
units of $N$) of the 10 fortunate traders and (in
the inset) the variation of the most probable and
average values of the residence time. (b) The
distribution of return-time to fortune (become
one of the 10 richest, starting from the 20th rank)
and (in the inset) the variation of the most probable
and average values of the return-time (in units of $N$). The error estimation is based on 10 runs. The
typical errors in the distribution of
both the residence and return times
are seen to grow with $N$ near the
most-probable values of the respective
quantities and are  indicated for $N$ =
400 here, when they are bigger than the
symbol sizes.
}
\label{fig:4b}
\end{figure}

For the wealth condensation in B-model
(with $f$ = 0), we show next in Fig. \ref{fig:4b}(a), the
distribution  of residence-time (in unit of $N$)
of the 10 fortunate traders and (in the inset)
the variation of the most probable and average
values of the residence-time ($\tau$, in unit of $N$).
For the same model
with $f$ = 0, we show in Fig. \ref{fig:4b}(b) the distribution of return-time
to fortune (become one of the 10 richest starting from the 20th rank)
and (in the inset) the variation of the most probable and average
values of the residence-time with market size $N$.

\subsection{Chakraborti or Yard-Sale model results}
\begin{figure}[!tbh]
    \centering
    \includegraphics[scale=0.75]{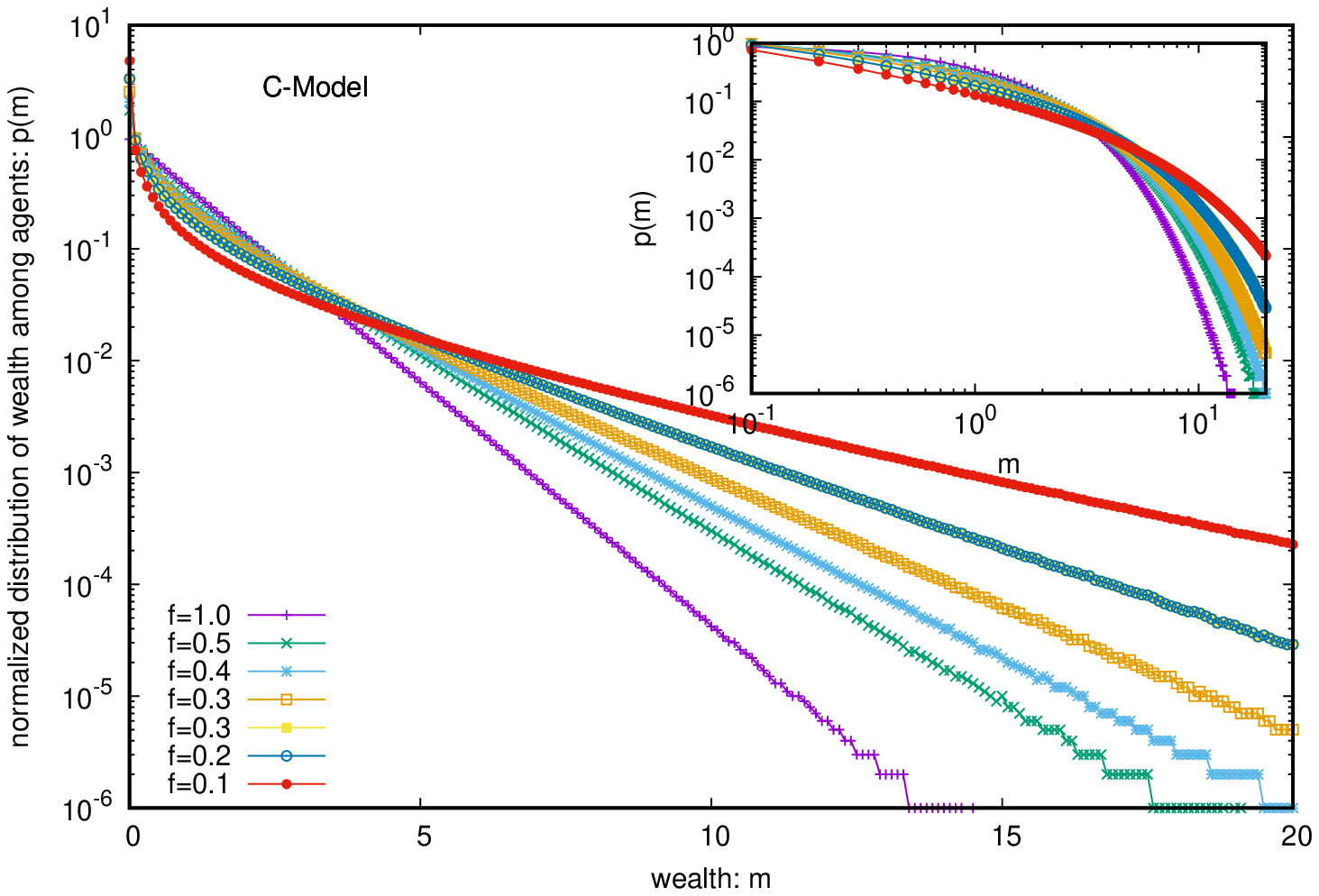}
    \caption{Wealth distribution $P(m)$ among all the agents against the
wealth $m$ in the C model for different probability  $f$ of DY random
exchanges. Note that the fluctuations appear to grow
more for the  lower values of the distribution
due to the log scale used in the y-axis.}
    \label{fig:5}
\end{figure}

\begin{figure}[!tbh]
    \centering
    \includegraphics[scale=0.75]{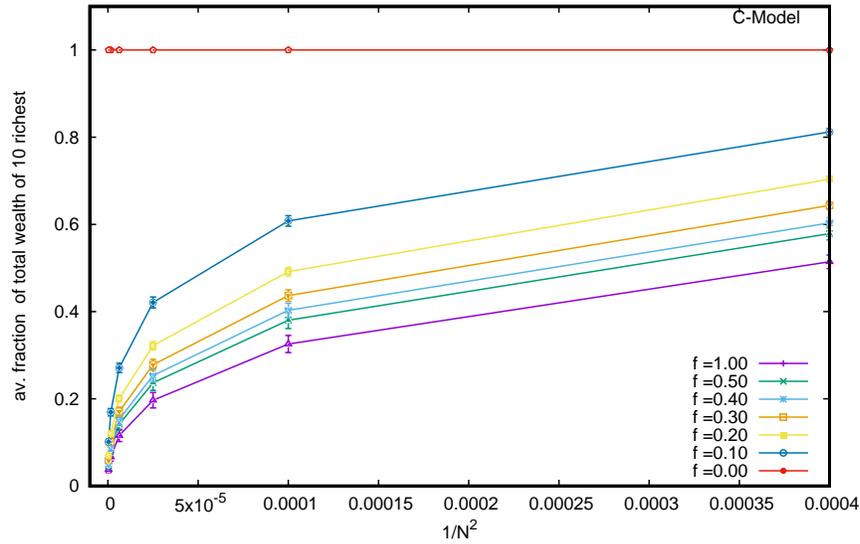}
    \caption{The limiting values (for large $N$) of the average fraction of
total wealth ($M = N$) possessed by the ten richest traders in the
the steady state of the C model with $f$ fraction of DY-like trades. For $f =
0$, the entire money goes to one agent and the other nine agents
contribute nothing. When we plot the fraction against $1/N^2$ (as with
DY type trades each of N traders interacts with $(N - 1)$ other
traders), the extrapolated values  all seem to approach
zero for any non-zero value of $f$. The error estimation is based on 10 runs. Typical sizes of error bars are indicated.}
    \label{fig:6}
\end{figure}

The C Model or Yard sale model is well-studied. However, in order to
check the stability of the condensation of wealth (entire money $M =
N$ going to the hand of one trader only, we added a nonvanishing
probability $f$ of each trader to follow DY trades or exchanges. We
see, immediately, the wealth condensation disappears for any $f >  0$
(see Fig. \ref{fig:5}) and the wealth gets distributed in the Boltzmann  form
(exponentially decaying with increasing wealth) among all the agents.
The inset shows that for any nonzero value of $f$, the steady state
wealth  distribution is exponentially decaying (and there is a power law
region) in this extended model C.  Also, when we consider the limiting
values (for large $N$) of the average fraction of total wealth ($M =
N$) possessed by the ten richest traders the steady state (see Fig. \ref{fig:6}), they all seem to vanish from the unit value  in the original C model
(with $f = 0$) for any non-zero value of $f$.

\subsection{Goswami-Sen model results}
\begin{figure}[!tbh]
    \centering
    \includegraphics[scale=0.75]{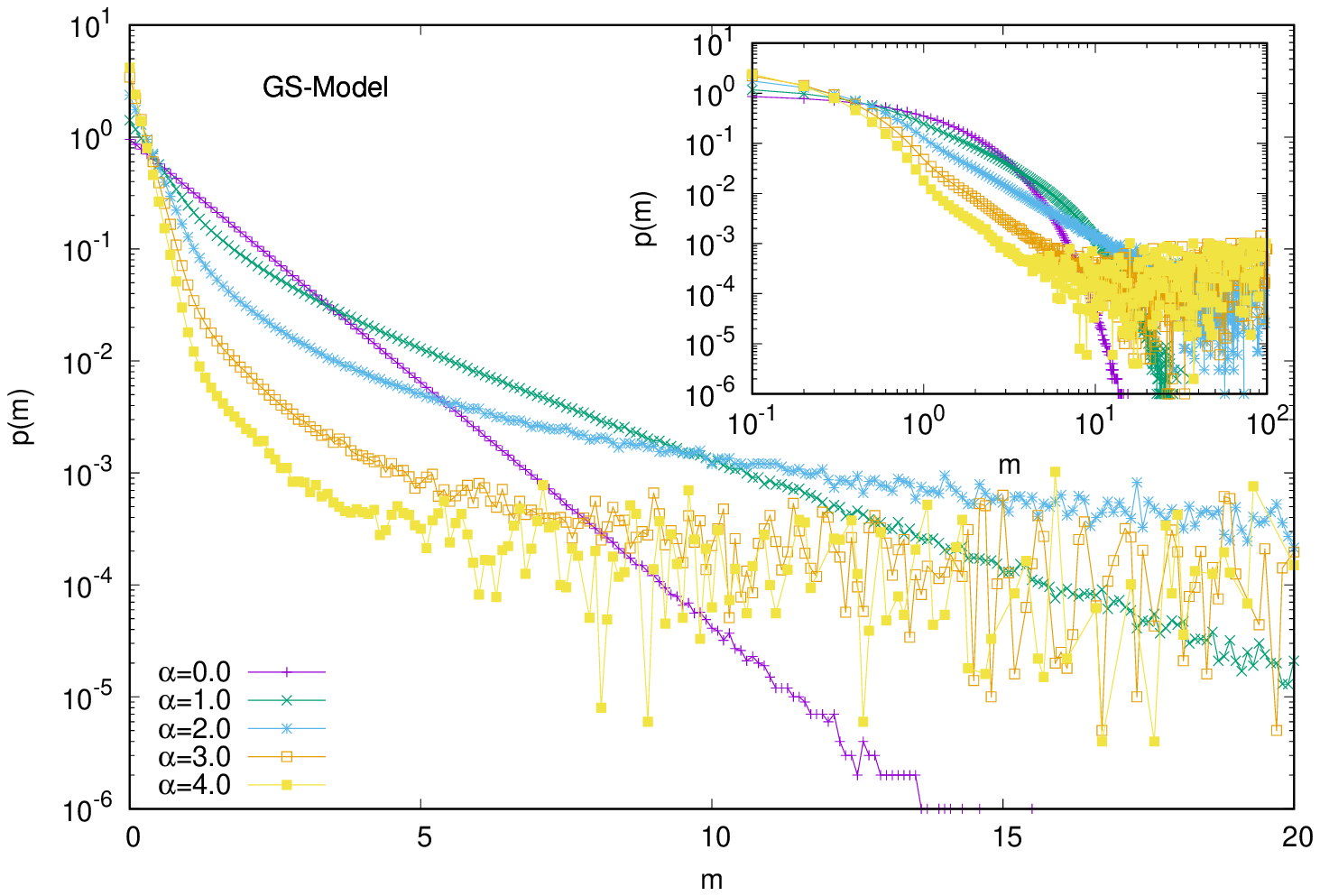}
    \caption{Wealth distribution $P(m)$ among all the agents
against the wealth $m$ in the GS model for different values of
$\alpha$. Note that the fluctuations appear to grow
more for the  lower values of the distribution
due to the log scale used in the y-axis.}
    \label{fig:7}
\end{figure}
\begin{figure}[!tbh]
    \centering
    \includegraphics[scale=0.75]{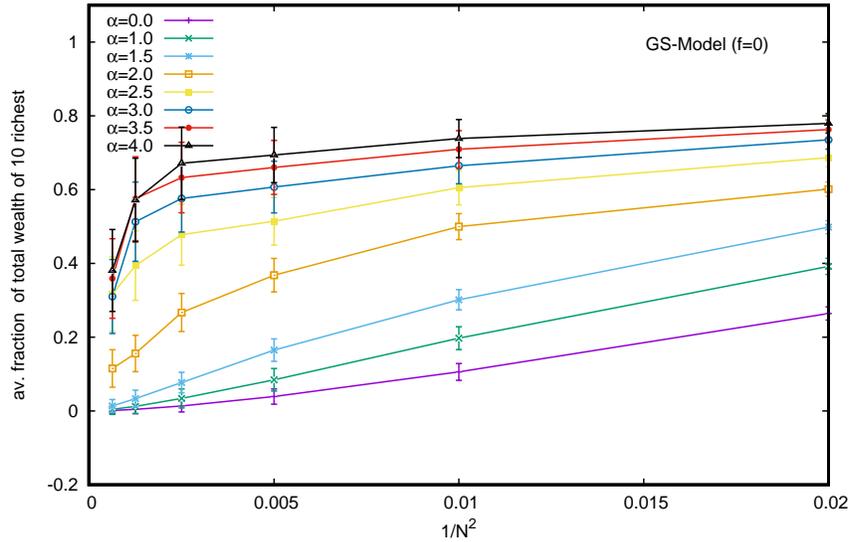}
    \caption{Plot of the fraction of total wealth ($M = N$) against
$1/N^2$ for different  values of $\alpha$ in the GS model. The
extrapolated (with $N$) values of the fraction  all seem to approach
zero for any non-zero value of $\alpha$. The error estimation is based on 10 runs. Typical sizes of error bars are indicated.}
    \label{fig:8}
\end{figure}

Here the interaction (trade) probability among the trade partners
($i$ and $j$) decreases with their wealth difference ($|mi – mj|$) at
that instant of trading (time), following a power law ($|mi –
mj|^{-\alpha}$). As such in the GS model, there is always a finite
(but small) probability of random exchanges.  We do not need to
consider the additional fraction of DY interaction in this model. Of
course, for $\alpha$ = 0, the model reduces to that of DY.  Our
numerical results confirm (see Fig. \ref{fig:7}) that  for $\alpha$ values less
then about 2.0, the steady-state wealth distribution among the traders
are still DY-like (exponentially
decaying). For higher values (beyond 2.0) of $\alpha$, power-law (Parto-like)
decays  occur (but no condensation of wealth).  Though the model leads to extreme  inequality, there is no condensation of wealth in the
hands of a few traders for any (larger) value of $\alpha$. In order
to check that we studied
again the  average fraction of total wealth ($M = N$) possessed by the
ten richest traders in the steady state of the GS model with $\alpha$.
When we plot the fraction against $1/N^2$  (see Fig. \ref{fig:8}), the
extrapolated values  of the fraction all seem to approach zero for any
 non-zero value for any of the $\alpha$ values considered.

\begin{figure}[!tbh]
    \centering
    \includegraphics[scale=0.55]{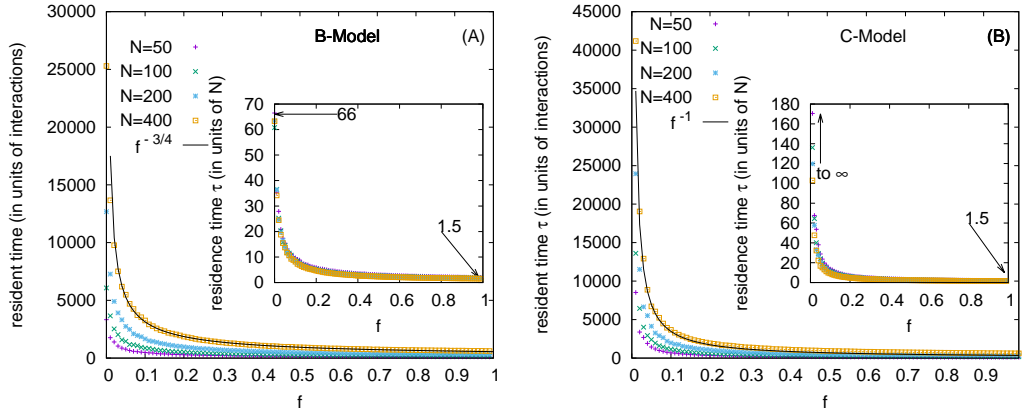}
    \caption{The DY fraction $f$ dependence of the bare
residence time (in units of interactions or exchanges)
at different $N$ values for B-model (Fig. 10a) and for
C-model (Fig. 10b) are shown. Their power law
fits with $f$ for $N$ = 400 are shown (for other $N$
vales, the respective prefactors change linearly with
$N$). The insets show the $f$ dependence of the
residence-time $\tau$ (in units of $N$). Note that the
limiting values of $\tau$ at $f$ = 0 is about 66 for
the B-model, while it goes to infinity for the C-model.}
    \label{fig:10}
\end{figure}

\section{SUMMARY and DISCUSSION}
In view of the observed extreme income or wealth
inequalities in society, the suitability of
the kinetic exchange models \cite{Chakrabarti2013} to  capture them,
at least qualitatively, have been investigated
here. We  distinguish between two types of such
extreme inequalities: One (Pareto) type \cite{Banerjee2023} where
a small fraction (typically 13\%) of the population
possess about 87\% of the total wealth   (following a power law distribution) of the
respective country. The other more recently
observed (and reported by  Oxfam \cite{Oxfam2020}) truly
extreme nature of income and wealth inequalities
worldwide, where only a handful number (say a few
hundreds to thousands) of super-rich people of
the world accumulate more than the total wealth of 50
to 60 percent poor people.

Several kinetic exchange models (see e.g, \cite{Chatterjee2004a,Chakrabarti2013})
have been developed to analyze Pareto type of
inequalities. We have investigated here  the statistics of some  kinetic exchange
models where, even in the $N$ going to infinity limit, only one person
can grab the entire wealth (as in the Yard-sale or Chakraborti or C model \cite{Chakraborti2002,Hayes2002}), or only 10 people can accumulate
about 99.98\% of the total wealth (as in the Banerjee or B model \cite{Banerjee2021},
see Fig. \ref{fig:2}). We investigate how these extreme inequalities in these
kinetic models get softened to the Dragulescu-Yakovenko (DY) \cite{Dragulescu2000} type
exponentially decaying wealth distributions among all the traders or
agents, when the traders each have a non-vanishing probability $f$ of
DY-type random exchanges. These condensations of wealth (100\% in the
hand of one agent in the C model \cite{Chakraborti2002}, or about 99.98\% in the hands of ten
agents in the B  model) then disappear in the large $N$ limit (clearly
seen when extrapolated against $1/N^2$, as in DY type random
exchanges, each of $N$ agents interact or exchange with all  others;
see Figs. \ref{fig:4} and \ref{fig:6}). We also showed that due to the built-in
possibility of DY-type random exchange dynamics in the Goswami-Sen or
GS model \cite{Goswami2014}, where the exchange probability decreases with an
inverse power of the wealth difference of the pair of traders, one
does  not see any wealth condensation phenomena. In both GS and B
model (with $f > 0$ fraction DY interactions or exchanges) no wealth condensation occurs,
though  strong Pareto-type power-law wealth distribution $P(m)$ or
inequalities occur  for large values of $\alpha$ and smaller values of
$f$ in GS and B models respectively (see Figs. \ref{fig:3} and \ref{fig:7}). For the wealth condensation in the B model, for $f$ = 0,
we additionally find that the fortunate top ten traders
are not unique and their fortune does not  last for 
long (residence-time $\tau$ to fortune  on
average is about 66 time units with its most
probable value around 25 time units, when counted
in units of $N$ trades or exchanges; see Fig. \ref{fig:4b}a).
The most probable `return-time' to such a fortune
(of the 20th rank holder to come within the group
of fortunate 10), is found to be about 20 (again
in units of $N$; see Fig. \ref{fig:4b}b). It may be noted that
with $f=0$, in the C-model, the residence-time $\tau$
is infinity for the only fortunate one accumulating
the entire wealth in the system. Indeed with increasing
values of DY fraction $f$, the values of $\tau$ in both
the cases decrease rapidly (see Fig. \ref{fig:10}), following inverse power
laws with $f$. We further note, for $f$ = 0
in the B-model near the most-probable values
of the wealth fractions (Figs. \ref{fig:1} and \ref{fig:2}) and
residence or return times (Fig. \ref{fig:4b}), the
fluctuations tend to grow with $N$,
indicating a possible divergence there in
the macroscopic limit of $N$. We plan to
explore its significance later.

Our  studies for the B, C, and GS kinetic exchange models, using Monte
Carlo techniques \cite{code}, suggest that the potential condensation type extreme
inequality can disappear in all of them if a non-vanishing probability
of random exchanges  are allowed, and converge to Pareto-type power
law inequality (for B and GS model) which in turn converges to Gibbs
like (exponentially decaying) wealth distribution for larger values of $f$  in the B model, smaller values of $\alpha$ in the GS model, or any values of $f>0$ in the C model.  These observations may  help to formulate public welfare policies.

\section*{acknowledgement}
SB acknowledges the support from DST INSPIRE. BKC is grateful to the Indian
National Science Academy for their Senior Scientist
Research Grant support.

\end{document}